\documentclass[prl,twocolumn,amsmath,amssymb,floatfix]{revtex4-1}
\usepackage{amssymb}
\usepackage{amsmath}
\usepackage{graphicx}
\usepackage{epsfig}
\usepackage{hyperref}
\hypersetup{colorlinks=true}

\usepackage[normalem]{ulem}

\usepackage{bm}% bs
\newcommand{\be}{\begin{equation}}
\newcommand{\eq}{\end{equation}}

\newcommand{\beq}{\begin{eqnarray}}
\newcommand{\eeq}{\end{eqnarray}}

\def\eq#1{(\ref{#1})}
\def\H1{\widehat{H}_1}

\begin{document}

\title[]{Something interacting and solvable in 1d}

\author{Eyzo Stouten$^{1,2}$, Pieter W. Claeys$^{1,3,4}$, Mikhail Zvonarev$^5$, Jean-S\'{e}bastien Caux$^{1}$, Vladimir Gritsev$^{1,5,6}$ }

\address{$^1$Institute for Theoretical Physics, Universiteit van Amsterdam, Science Park 904,
Postbus 94485, 1098 XH Amsterdam, The Netherlands\\
$^2$Fachbereich Physik, Bergische Universit\"{a}t Wuppertal, 42097 Wuppertal, Germany\\
$^3$Department of Physics and Astronomy, Ghent University, Krijgslaan 281-S9, 9000 Ghent, Belgium\\
$^4$Center for Molecular Modeling, Ghent University, Technologiepark 903, 9052 Ghent, Belgium \\
$^5$LPTMS, CNRS, Univ. Paris-Sud, Universit\'{e} Paris-Saclay, 91405 Orsay, France\\
$^6$Russian Quantum Center, Skolkovo, Moscow 143025, Russia}
%\ead{V.Gritsev@uva.nl}
\begin{abstract}
We present a two-parameter family of exactly solvable quantum many-body systems in one spatial dimension containing the Lieb-Liniger model of interacting bosons as a particular case. The principal building block of this construction is the previously-introduced \cite{SCCG} family of two-particle scattering matrices. We discuss an $SL(2)$ transformation connecting the models within this family and make a correspondence with generalized point interactions. The Bethe equations for the ground state are discussed with a special emphasis on ``non-interacting modes" connected by the modular subgroup of $SL(2)$. The bound state solutions are discussed and are conjectured to follow some correlated version of the string hypothesis.
The excitation spectrum of the new models in this family is derived in analogy to the Lieb-Liniger model and we show that for certain choices of parameters a spectrum inversion occurs such that the Umklapp solutions become the new ground state.

\end{abstract}

\maketitle

%\section{Introduction}
\section{Introduction}
Exactly solvable models play a profound role in our understanding of low-dimensional statistical mechanics and condensed matter physics. Of particular importance is the Lieb-Liniger (LL) model, representing bosons in one spatial dimension interacting through a $\delta$-function potential of arbitrary strength $c$,
\beq
H_{LL}=-\sum_{j=1}^{N}\frac{\partial^{2}}{\partial x_{j}^{2}}+c\sum_{i\neq j}^N \delta(x_{i}-x_{j}).
\eeq

Originally employed to extend the Bogoliubov theory predictions for the excitation spectrum~\cite{LL1, LL2}, it has emerged in quantum optics for modeling effective Kerr nonlinear media in the quantum regime \cite{Lai1,Lai2,Yudson}, and photonic many-body correlated states \cite{Chang,Imamoglu}. In two-dimensional statistical mechanics this model appeared in the description of an interface interacting with impurities~\cite{Kardar}, establishing link of this problem to the Kardar-Parisi-Zhang universality class~\cite{KPZ,CD,Dotsenko}. This could then also be connected to the asymmetric simple exclusion process, as discussed in Ref.~\cite{TW}. Experiments with ultracold atomic systems can nowadays confine the motion of the constituent particles to one spatial dimension (1D), if the two-particle scattering is determined by an interaction of the $\delta$-function form, then LL model emerges~\cite{Olshanii_98}. This model then describes experimentally observed phenomena such as interaction-induced fermionization of bosons~\cite{KWW1,KWW2,Bloch}, suppression of three-body recombination~\cite{Tolra_04, Haller_11}, absence of thermalization~\cite{Kinoshita_06}, stability in the attractive regime~\cite{Haller_09}, and Bloch oscillations in the absence of a lattice~\cite{Meinert_17}. For a more detailed review we refer the reader to Ref.~\cite{rev}.

However, the LL model does not exhaust all integrable continuum models in 1D.
Gaudin~\cite{Gaudin1}, Yang~\cite{Yang1,Yang2}, and Sutherland~\cite{Sutherland1} demonstrated that the solution to the LL model can be further extended to a problem of $\delta$-interacting particles with no limitation on the symmetry of the wave function. In particular, the Yang-Gaudin model is a generalization to interacting fermions, and multicomponent mixtures of point-interacting models have also been found by Sutherland. In the remarkable paper \cite{Yang1}, C. N. Yang then generalized the LL model to particles in arbitrary representations of the symmetric group.

The key ingredient of integrability in 1D is the Yang-Baxter equation (YBE) for the two-particle scattering matrices $\check{S}_{ij}(k_i,k_j)$ of two particles labeled by $i$ and $j$, given by
% %%
% \beq
% &\check{S}_{ij}(k_{i},k_{j})\check{S}_{ik}(k_{i},k_{k})\check{S}_{jk}(k_{j},k_{k}) \qquad \qquad \nonumber \\
% &\qquad \qquad  =\check{S}_{jk}(k_j,k_k)\check{S}_{ik}(k_i,k_k)\check{S}_{ij}(k_i,k_j),
% \eeq
% %%
%%
\beq
&\check{S}_{jk}(k_{i},k_{j})\check{S}_{ij}(k_{i},k_{k})\check{S}_{jk}(k_{j},k_{k}) \qquad \qquad \nonumber \\
&\qquad \qquad  =\check{S}_{ij}(k_j,k_k)\check{S}_{jk}(k_i,k_k)\check{S}_{ij}(k_i,k_j),
\eeq
in which $k$ is the scattering rapidity of a particle. The scattering matrix found by Yang \cite{Yang1} was expressed in terms of the permutation operator $\Pi_{ij}$, which interchanges quantum spaces of two interacting particles, reading as
\beq
\check{S}_{ij}(k_{i},k_{j})=\frac{\Pi_{ij}-iF(k_{i},k_{j})}{\openone+iF(k_{i},k_{j})}.
\label{S-matrix}
\eeq
Using Artin's braid relation for the permutation operator, $\Pi_{12}\Pi_{23}\Pi_{12}=\Pi_{23}\Pi_{12}\Pi_{23}$, combined with the fact that ${\Pi_{ab}}^{2}=\openone$, one can easily see that the function $F_{ij}\equiv F(k_{i},k_{j})$ has to satisfy the functional equation
\beq
F_{ij}F_{ik}+F_{ik}F_{jk}=F_{ij}F_{jk}.
\label{F-eq}
\eeq
C. N. Yang found a {\it particular} solution to this equation,
\beq
F^{Yang}_{ij}=\frac{c_{0}^2}{k_{i}-k_{j}},
\label{F-Yang}
\eeq
where $c_{0}$ is a free parameter. By specifying a particular representation of the symmetric group generated by $\Pi_{ij}$ one can generate a large class of known continuum models. The LL model with $c=c_0^2$ follows from the totally symmetric representation, while the model of interacting spin-1/2 particles can be obtained by taking $\Pi_{ij}=\frac{1}{2}(1+\vec{\sigma}_{i}\cdot\vec{\sigma}_{j})$ with $\vec{\sigma}=(\sigma_x,\sigma_y,\sigma_z)$ the vector composed of the three Pauli matrices.

The purpose of this paper is to generalize the solution of Lieb and Liniger. In Ref. \cite{SCCG} we found a {\it new solution} to Eq.~(\ref{F-eq}) in the class of {\it rational } functions. This solution is parametrized by {\it two} free (possibly complex) interaction parameters and thus generates a new class of integrable many-body 1D systems with possibly as many physical applications as the LL model. The symmetry properties and some consequences of this solution are discussed here. Numerical solutions of the resulting Bethe ansatz equations are presented and several important limits are analyzed. We conclude that the model introduces many surprises which await further analysis.

\section{Generalization of the model of interacting bosons}
The solution of any many-body problem starts with the analysis of the two-body case. A two-parametric generalization of the two-body scattering matrix from Eq. (\ref{F-Yang}) was presented in Ref.~\cite{SCCG}, reading as
\beq
F_{ij}=\frac{c_{0}^{2}+c_{0}c_{1}(k_{i}+k_{j})+c_{1}^{2}k_{i}k_{j}}{ k_{i}-k_{j}},
\label{F-new}
\eeq
where both $c_{0}$ and $c_{1}$ are {\it free} parameters.

This two-parameter solution has two clear limiting regimes. Yang's scattering matrix is reproduced in the $c_1=0$ case, leading to the bosonic LL model with the usual $\delta(x)$-interactions, whereas the $c_0=0$ case returns the scattering matrix used by Cheon-Shigehara (CS) \cite{CS-model,CS-potential}, representing a fermionic model with $\delta''(x)$ contact interactions. Furthermore, in the limit when only the first two terms in the numerator of $F_{ij}$ are kept, namely when $c_0 \gg c_1$, one should retrieve results similar to those found in Refs.~\cite{BGH,GBK,Kundu}, where the connection with anyonic models is studied.

%%%%%%%%%%%%%%%%%%%%%%%%%%%%%%%%
\subsection{$SL(2)$ group structure of the model}
In this section we repeat some of our previous findings \cite{SCCG} and specialize them to continuum models. Previously, we noted that our solution has an intrinsic $SL(2)$ duality symmetry associated with a transformation of the rapidities combined with a transformation of the couplings $c_{0,1}$. One can notice that if we transform the rapidities $k_{i}$ and $k_{j}$ of the scattering matrix (\ref{S-matrix}) and $F_{ij}$ from Eq.~(\ref{F-new}) according to the fractional-linear (M\"{o}bius) transformation,
\beq
\tilde{k}_{i,j}=\frac{\alpha k_{i,j}+\beta}{\gamma k_{i,j}+\delta},\qquad \alpha\delta-\gamma\beta=1,
\eeq
the solution (\ref{F-new}) remains the same {\it iff} we simultaneously transform the couplings $c_{0}$ and $c_{1}$ as
\beq
\left(
\begin{array}{c}
  \tilde{c}_{0}   \\
 \tilde{c}_{1}
\end{array}
\right)=\left(
\begin{array}{cc}
  \delta   & \beta\\
 \gamma & \alpha
\end{array}
\right)\left(
\begin{array}{c}
  c_{0}   \\
 c_{1}
\end{array}
\right).
\eeq
Here, the unimodularity condition $ \alpha\delta-\gamma\beta=1$ is essential. The rapidities $k_{i,j}$ are in principle allowed to take arbitrary complex values (corresponding to bound states), so in principle the parameters $\alpha,\beta, \gamma,\delta$ could be complex as well, thus transforming under the group $SL_{k}(2,\mathbb{C})$ (the subscript $k$ denotes that they act on rapidities). At the moment, the couplings $c_{0,1}$ can be considered as complex as well.  The $SL(2,\mathbb{R})$ transformation connects rapidities of {\it different} models with different coupling constants, and so it acts as a {\it duality} symmetry, similar to the electric-magnetic duality in field theory. In particular, the LL and CS models are connected by the inversion
\beq
k_{j}\leftrightarrow -k_{j}^{-1},
\eeq
which would correspond to the ${\cal S}$-transformation in the modular group case (see below).

Interestingly, one can understand the structure of the model by looking into the algebraic properties of $SL(2)$. Indeed, one can notice that the vector $(c_{0},0)^{T}$ (corresponding to the LL model) only undergoes a scaling when transformed by the upper-triangular matrices. The vector $(0,c_{1})^{T}$ (corresponding to the fermionic model of CS) has the same property with respect to the lower-triangular matrices. Algebraically speaking, upper-triangular matrices represent the parabolic elements of $SL(2)$. In general, $SL(2)$ has four {\it conjugacy classes} characterized by the trace  $(\alpha+\delta)$:  the parabolic class, when $\alpha+\delta=\pm 2$; the elliptic class, if $|\alpha+\delta|\leq 2$ and $\alpha+\delta$ is real; the hyperbolic class if $\alpha+\delta$ is real and $|\alpha+\delta|\geq 2$, and finally the loxodromic one, when $\alpha+\delta$ is a complex number. Geometrically, these different classes correspond to different transformations: the parabolic class (generated by the nilpotent element of the algebra) is responsible for the {\it shear mappings} (the LL model), while the elliptic elements are interpreted as Euclidean rotations, while the hyperbolic elements correspond to squeeze mappings (boosts) of the plane.

A particularly interesting subgroup of $SL(2)$ is given by $SL(2,\mathbb{Z})$, the group of matrices with integer elements and unit determinant. Indeed, we will show in the next section that this (modular) subgroup plays an important role in physics of the model.

The group-theoretical aspect of the model brings the following parallel with a problem of self-adjoint extension \cite{Seba,Albeverio,rev-VS,Bajnok}.
Consider a single particle problem, where the kinetic energy operator is defined as $K=d^{2}/dx^{2}$. The requirement of self-adjointness  $\langle \psi K|\psi\rangle=\langle\psi| K\psi\rangle$ in the domain $\Omega=R-\{0\}$ as written in coordinate representation translates into
\beq
0 &= &\int_{\Omega}dx[\psi^{*}K\psi-(K\psi)^{*}\psi] \\
&=& [\psi^{*}\psi'-(\psi')^{*}\psi](0^{+})-[\psi^{*}\psi'-(\psi')^{*}\psi](0^{-}), \nonumber
\eeq
where we have applied integration by parts. This condition is equivalent to the requirement of probability current $j$ conservation across the boundary at $x=0$, where $j=-i[(\psi')^{*}\psi -\psi^{*}\psi']/2$. Considering gluing conditions consistent with self-adjoint extensions at $x=0$ for a function $\psi$ and its derivative $\psi'$, it was shown in Ref. \cite{Seba} (see also Section 1 of Ref. \cite{Bajnok} for a review) that the most general gluing condition is given by
\beq
\left(
\begin{array}{c}
  \psi(0^{+})   \\
 \psi'(0^{+})
\end{array}
\right)=e^{i\phi}\left(
\begin{array}{cc}
  a   & b \\
 c & d
\end{array}
\right)\left(
\begin{array}{c}
  \psi(0^{-})   \\
 \psi'(0^{-})
\end{array}
\right),
\label{transfer-mat}
\eeq
where $\phi\in[0,\pi]$ and the real coefficients $a, b, c, d$ satisfy the unimodularity condition
\beq
ad-bc=1.
\eeq
The central point of this construction is the conservation of the {\it probability current } across the internal boundary. This can be compared with the case of the 1D Bose gas, where, because of the bosonic symmetry of the wave function, regularity of the wave function has to be imposed, while its spatial derivative can experience a jump proportional to the interaction strength. In the case of fermions the opposite situation occurs -- the derivative is continuous while the wave function exhibits a jump discontinuity. The most general potential $v(x)$ respecting these internal jump boundary conditions is a combination of $\delta(x)$, $\delta'(x)$ and $\delta''(x)$ \cite{Seba,Albeverio,rev-VS,Bajnok},
\beq
\frac{1}{2}v(x)&=&g_{1}\delta(x)-(g_{2}-ig_{3})\frac{d}{dx}\nonumber\\
&+&(g_{2}+ig_{3})\frac{d}{dx}\delta(x)-g_{4}\frac{d}{dx}\delta(x)\frac{d}{dx},\label{V}
\eeq
where the parameters $g_{i},\ i=1,\ldots, 4$ can be related to the elements of the transfer matrix (\ref{transfer-mat}) across the boundary at $x=0$. Note that the bosonic and fermionic limits nicely correspond to the lower- and upper-triangular matrices, similar to the discussion of $SL_{c}(2) $ in the scattering matrix (note the transposition involved in going from $SL_{k}(2)$ to $SL_{c}(2)$).

Combining this with explicit computations similar to Ref.~\cite{BGH}, motivates us to {\bf conjecture} \cite{foot1} that the scattering matrix corresponding to $F_{ij}$ in Eq. (\ref{F-new}) gives rise to the following many-body integrable Hamiltonian
\beq
H_{F} = -\sum_{j=1}^N \frac{\partial^{2}}{\partial x_{j}^{2}} +\sum_{j<k}^N&& \left(c_{0}-ic_1\frac{\partial}{\partial x_{j}}\right)\left(c_{0}-ic_1\frac{\partial}{\partial x_{k}}\right) \nonumber \\
&& \qquad \qquad \times     \delta(x_{j}-x_{k}),\label{2parH}
\label{HF}
\eeq
encompassing and extending both the LL and SC models.

It is known that the LL model corresponds to the nonlinear Schr\"odinger equation in the field theoretic formulation \cite{Bajnok}. Inspired by this we propose that our Hamiltonian can be interpreted as a model which (a) incorporates a density-density interaction $c_0^2 \rho(x)\rho(x)$, a density-current interaction $c_0c_1\rho(x)j(x)$ and a current-current interaction $c_1^2j(x)j(x)$, where $\rho=\Psi^\dag(x)\Psi(x)$ and $j(x)=\Psi^\dag(x)\partial_x\Psi(x)$. In these terms, our duality collects the densities and currents into a unified object which transforms under the $SL(2)$ group described above.

%In other words we assume that the wave function can be written down and the Hamiltonian can be derived along the lines of Batchelor, Guan and Kundu \cite{GBK}, which is how this conjecture was obtained. However as of now we have not had the time to redo this calculation because we are not completely convinced of its correctness in the light of the publications by Patu, Korepin and Averin \cite{PKA}, therefore we stated it as a conjecture and hope to comment on it in later times.}.

%%%%%%%%%%%%%%%%%%%%%%%%%%%%%%%%%%%%%
\section{Bethe ansatz equations and roots}
%%%%%%%%%%%%%%%%%%%%%%%%%%%%%%%%%%%%%
\subsection{Bethe equations and non-interacting modes}
While the $SL(2)$ symmetry is preserved on the level of the scattering matrix and the YBE, it acts nontrivially on a system with fixed external boundary conditions, in that sense relating different physical models. Let's demonstrate this explicitly by constructing Bethe equations for our new solution (\ref{F-new}) following the standard procedure. Namely, we impose periodic boundary conditions and move the particle $j$ around a ring of size $L$. Every time this particle experiences a collision with another particle (say $i$) a scattering phase shift $\check{S}_{ij}(k_{i},k_{j})$ is collected. Moving the particle completely around the ring of $N$ particles, the periodic boundary conditions impose a single-valued condition on the wave function which results in the consistency equations
\beq
e^{ik_{j}L}\prod_{i\neq j}^{N}\check{S}^{F}(k_{j},k_{i})=1,
\label{g-BA-eq}
\eeq
where $\check{S}^{F}(k_{i},k_{j})$ is defined as in Eq.~(\ref{S-matrix}) with $F$ given by Eq.~(\ref{F-new}). It is important to note that here we {\bf assume} that there is a basis in which the permutation operator which enters Eq. (\ref{S-matrix}) can be chosen to be proportional to identity operator: $\Pi_{ij}=\openone$ \cite{foot2}.
In this case Eqs.~(\ref{g-BA-eq}) serve as generalized Bethe ansatz equations (BAE) which define the allowed values of the momenta $k_{j}$.

Remarkably, it follows directly from Eqs.~(\ref{S-matrix}) and (\ref{F-new}) that there are special points in the space of rapidities when the scattering matrix trivializes, $\check{S}^{F}(k_{i},k_{j})\equiv \openone$. This happens when either one of the rapidities  satisfies
\beq
k_{j}= -\frac{c_{0}}{c_{1}}. \label{singpnt}
\eeq
In this case Eq.~(\ref{g-BA-eq}) implies $\exp(ik_{j}L)=1$, which means that at these points there are {\it free}, non-interacting modes in a system with $k_{j}=-c_0/c_1=2\pi n_{j}/L$ for integer $n_{j} \in \mathbb{N}$. This quantization of couplings (in units of $L$) combined with the $SL(2,\mathbb{R})$ duality implies that a subgroup of the latter, the {\it modular} group $SL(2, \mathbb{Z})$ connects, in a sense of duality transformations, all free, noninteracting modes of the model. In the Yang model, the scattering function $F_{ij}$ could vanish only if one of the momenta $k_i$ or $k_j$ becomes infinite. However, due to the symmetry in the new solution, this can happen at finite momenta as long as $c_1 \neq 0$. (Mapping Yang's solution for $k_{i,j} \to \infty$ and $c_1=0$ to finite $\tilde{k}_{i,j}$ always results in $\tilde{c}_1 \neq 0$.) This vanishing has major consequences in both the Lieb-Liniger and the related Richardson-Gaudin class of models as introduced in Ref.~\cite{SCCG}. Given an eigenstate in the Lieb-Liniger model, any particle with trivial scattering ($k = -c_0/c_1$) can be added and the resulting state will still satisfy the Bethe ansatz equations provided the quantization condition $k = \pm 2 \pi n/L, n \in \mathbb{N}$ is simultaneously satisfied. While this restricts the values of $c_0/c_1$ for which such particles can be added for finite systems, this will always be possible in the thermodynamic limit. Here, the Bethe ansatz equations can be recast as a single integral equation, which will exhibit a singularity following from this non-interacting mode (see Appendix).

\vspace{8pt}
\subsection{Ground state root distribution}
% \vspace{-25pt}
\begin{center}
\begin{figure}[t]
%\vspace{3pt}
\includegraphics[width=\columnwidth]{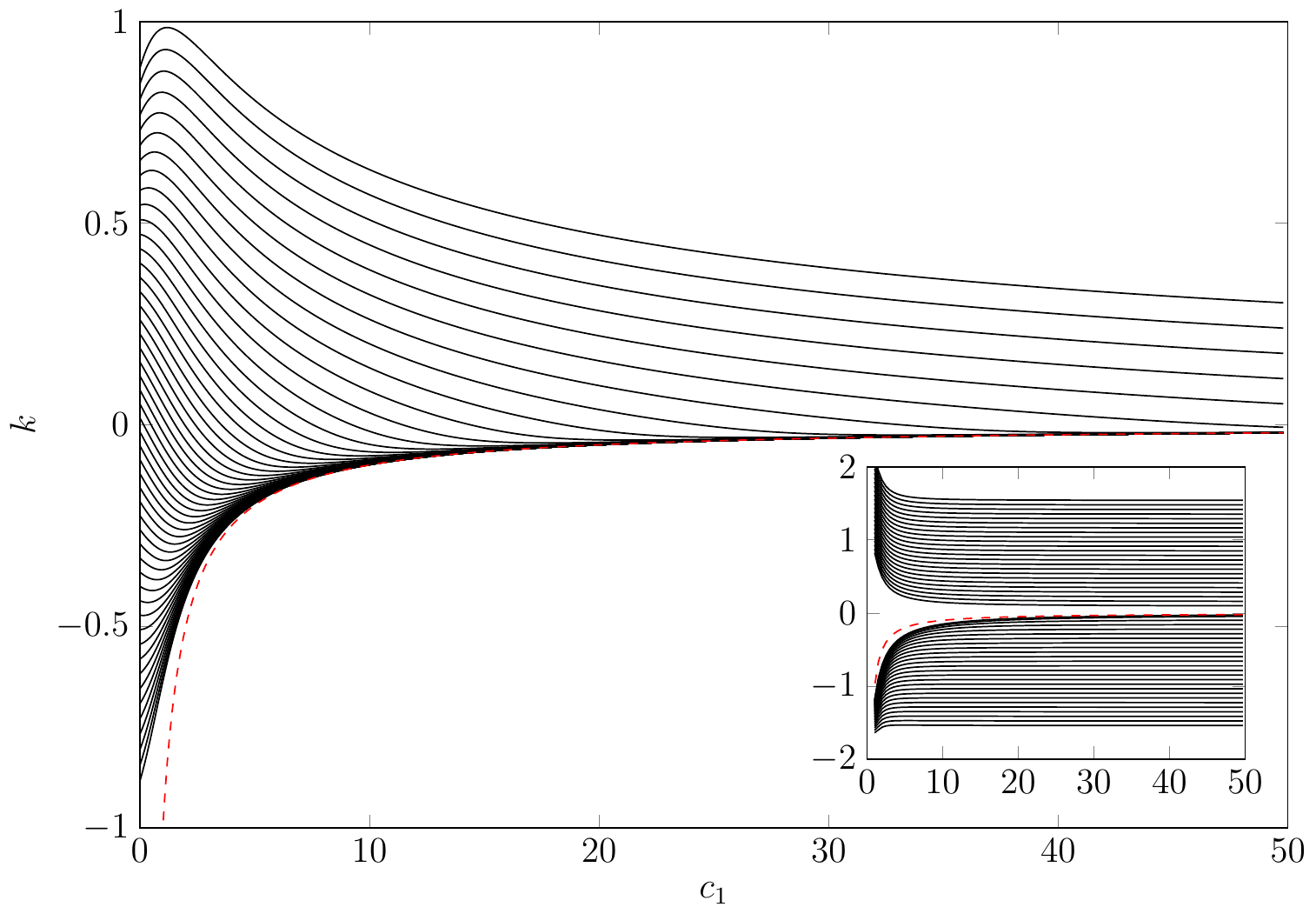}
\caption{Ground state rapidity distribution for different values of $ 0\leq c_1\leq 50$ for $c_0=1$, number of particles $N=50$ and system size $L=100$. The main plot corresponds to the rapidities obtained starting from the limit $c_1 \ll c_0$ while the evolution of rapidities starting from the opposite limit $c_1 \gg c_0$ is plotted in the inset. The line $k=-c_0/c_1$ is plotted in dashed red.}
\label{fig:BackwardsRoots}\vspace{3pt}
\includegraphics[width=\columnwidth]{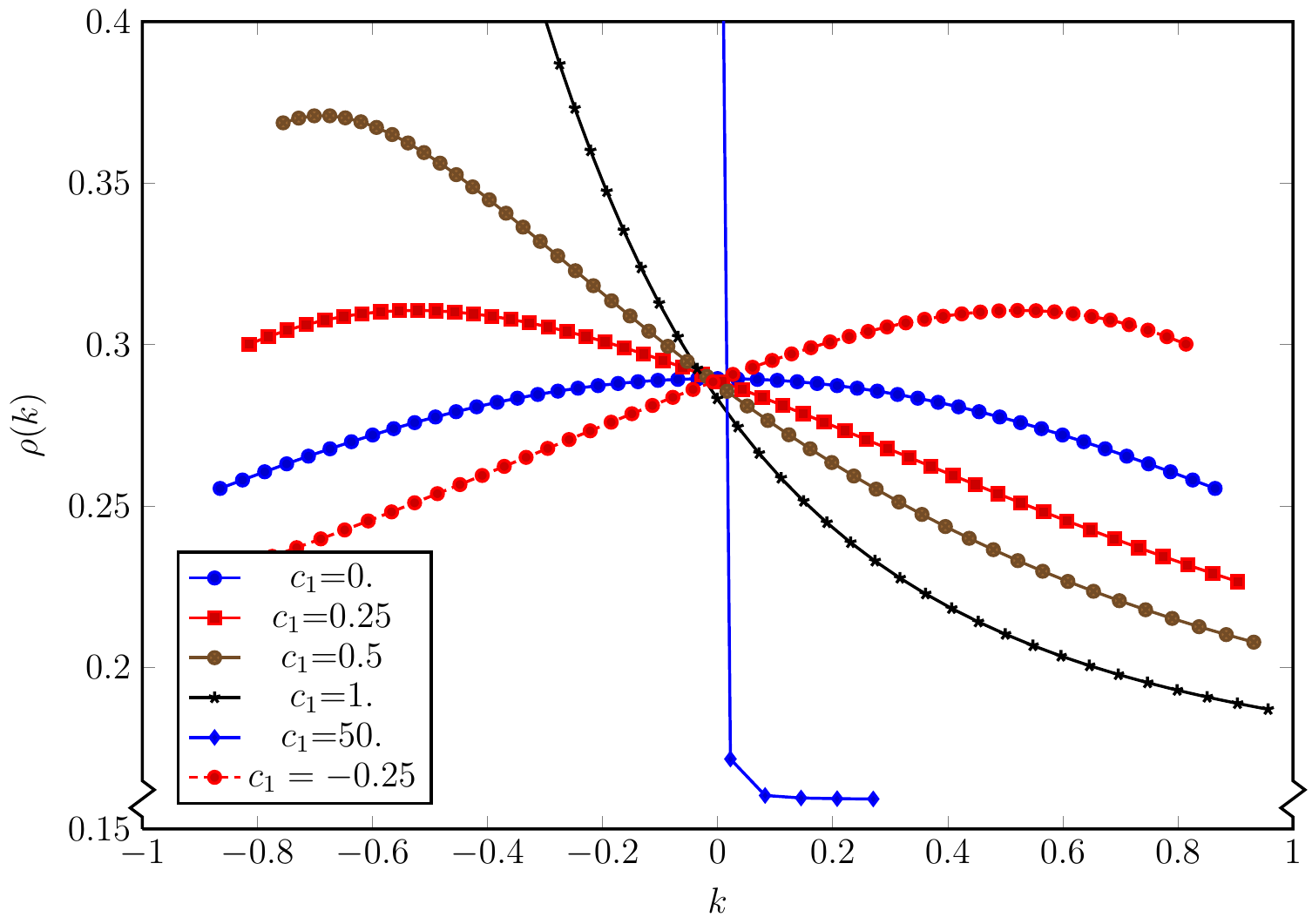}
\caption{Ground state density for different values of $ -0.25 \leq c_1\leq 50$, starting at $c_1 \ll c_0$. Here $c_0=1$, the number of particles is $N=50$, and the system size is $L=100$.}
\label{fig:Density}
\end{figure}
\end{center}
\vspace{-16pt}

In the following sections the numerical solutions to the BAE (\ref{g-BA-eq}) will be discussed. These solutions were obtained from the familiar logarithmic form of the BAE for real, fixed $c_0$ and varying $c_1$. In Fig.~\ref{fig:BackwardsRoots} the ground state solution of the BAE for different $c_1$ is given using as initial values $c_1 \ll c_0$ and $c_1 \gg c_0$, where the ground state is known, following from the correspondence to the LL and CS models. By ground state solution we mean the state following from the solution with (half) integer quantum numbers distributed around zero in these limiting cases. While the ground states are explicitly known in both limits, it can immediately be observed that the resulting solutions do not match. This implies a crossing between both ground states at some value of $c_1 \neq 0$, where the original solution will correspond to an excited state. This can be interpreted as a direct consequence of the non-interacting mode $k=-c_0/c_1$, since it is impossible for rapidities to cross this line as $c_1$ is changed. Indeed, if at some point $k=-c_0/c_1$, the resulting rapidity decouples from the system and remains a non-interacting mode as $c_1$ is varied (this can be easily seen in the thermodynamic limit). In the $c_1 \ll c_0$ limit the LL model is obtained, for which all rapidities at small values of $c_1$ lie at one side of the diverging line $-c_0/c_1$. This can be contrasted with the CS limit, where the rapidities are symmetrically distributed around $0$ for $c_1 \to \infty$, and hence half the rapidities remain above and half below the non-interacting line $k=-c_0/c_1$ as $c_1$ is varied. As such, it is impossible to continuously deform these solutions into each other, and a level crossing (or multiple level crossings) needs to occur when changing $c_1$ between the two known limits.

As further illustration, the root density
\beq
\rho\left(\frac{k_j+k_{j+1}}{2}\right)=\frac{1}{L(k_{j+1}-k_j)},
\eeq
is displayed in Figure \ref{fig:Density}, where the solution starting from the LL limit exhibits the characteristic semicircular root density distribution around zero for $c_1=0$ \cite{LL1}. Small positive and negative values of $c_1$ are also shown for comparison to the anyonic behavior at $c_1 \ll c_0$ \cite{BGH,GBK}.

Due to decoupling of the roots at $k_j=-c_0/c_1$ the distributions of $k_j$ convergence asymptotically to $-c_0/c_1$. This is confirmed by the study of the weakly-interacting Gaudin limit $c_0 \ll 1$ (see below). Starting from $c_1 \ll c_0$ (LL initial conditions, main plot Figure~\ref{fig:BackwardsRoots}), $c_1\rightarrow \infty$ also results in a condensate around $k_j=0$, which is surprising because in this limit the model should exhibit fermionic behavior as noted by CS \cite{CS-model}. Nevertheless, for large $c_1$ all roots have decoupled because they approach $k_j =-c_0/c_1$, which does happen in fermionic models and is not contradictory to the findings of CS.

The second initial value $c_1 \gg c_0$ can be considered a second Tonks-Girardeau limit in $c_1$, giving the same results as the single parameter case ($c_0 \gg 1$ and $c_1=0$), leading to $k_j = 2 \pi n/L$ where $n\in \mathbb{Z}$ (cf. Figure~\ref{fig:BackwardsRoots}). However, when approaching $c_1\sim 0$ this limit does not retrieve the same state as obtained from the first initial value. Instead the upper roots tend to infinity and the lower roots approach the line $k=-c_0/c_1$. At this point the roots decouple and numerical analysis becomes unstable, leaving this region an open problem. This unstable behavior is only present in states that contain roots the that hit the line $k_j=-c_0/c_1$. For highly exited states which have $k_j>-c_0/c_1$, $\forall j=1 \dots N$ and $\forall c_1\geq0$ the two initial states do connect, since then all rapidities lie at the same side of the non-interacting line.

The overall connection of the CS and LL initial states thus remains an open problem. We {\bf conjecture} that the solution seems to be related to states other than the ground state through the ``non-interacting modes'' due to the asymptotic decoupling and the excitation spectrum as discussed in a later section.

We note that there is no definite property of the scattering phase with respect to $k_{j}\rightarrow -k_{j}$ transformation. This is also manifested in both figures \ref{fig:BackwardsRoots} and \ref{fig:Density} that for $c_1 \neq 0$ the {\it parity symmetry} of the model is broken. We {\bf conjecture} the following picture: Physically, the ground state perhaps can be interpreted in terms of two counter-propagating liquids with different densities and velocities. The overall momentum of all the above discussed states for any $c_1$ however vanishes (as confirmed by numerics). Furthermore the domain for $c_1<0$ shows exactly the same behavior as the previously discussed cases for $c_1>0$ except for an inversion of the root spectrum in $k$ due to the line $-c_0/c_1$ now coming from above.

%%%%%%%%%%%%%%%%%%%%%%%%%%%%%%%%%%%%%
\subsection{Gaudin limit}
%%%%%%%%%%%%%%%%%%%%%%%%%%%%%%%%%%%%%
In order to better understand the asymptotic behaviour as the rapidities approach the line $-c_0/c_1$ for large $c_1$, the limit $c_0 \ll 1$ can be investigated, where the rapidities can be connected to the roots of orthogonal polynomials. In the LL model, it is known that for $c_0 \ll 1$ the rapidities can be obtained as roots of Hermite polynomials \cite{gaudin_bethe_2014}, and this can be extended to the current model for arbitrary $c_1$.

\begin{center}
\begin{figure}[h]
\includegraphics[width=\columnwidth]{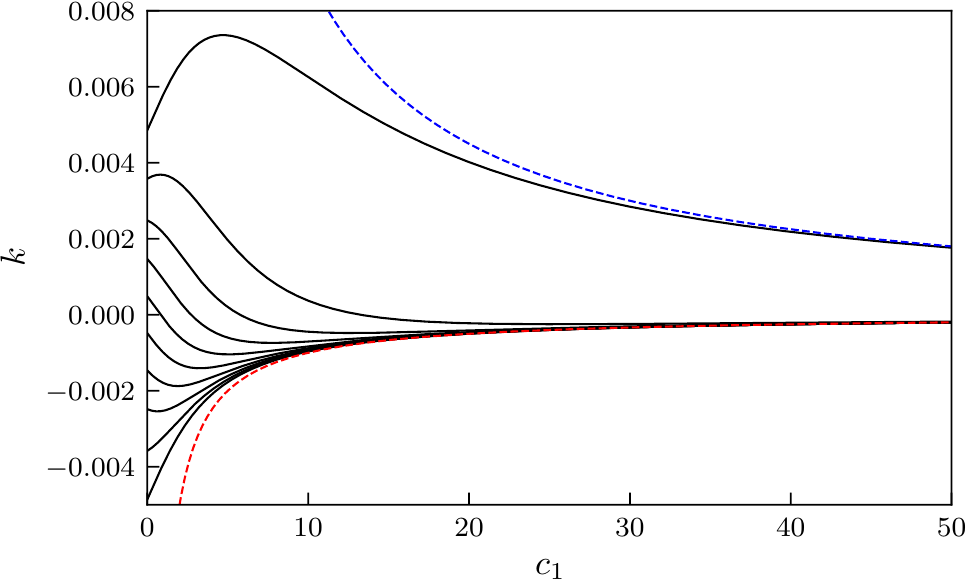}
\caption{Approximate rapidities from the Gaudin limit for $c_0=0.01$, $N=10$ and $L=100$ from the roots of associated Laguerre polynomials, with the dashed red line denoting $k=-c_0/c_1$ and the dashed blue line denoting $k=(N-1)c_0/c_1$. The qualitative behavior of the rapidities is the same as for larger values of $c_0$ starting from the LL limit.}
\label{GaudinLim}
\end{figure}
\end{center}

As shown in the Appendix, a series expansion for the rapidities can be obtained as
\beq
k_j = -\frac{c_0}{c_1}+\frac{c_0}{c_1^3}\frac{L}{z_j} + \mathcal{O}(c_0^2),
\eeq
with $z_j$ the $j$-th root of the associated Laguerre polynomial $L_N^{\alpha}(z)$ with $1+\alpha = L/c_1^2$ \cite{szego_orthogonal_1975}. These roots can be obtained for arbitrary values of $\alpha$ by solving the recursion relation satisfied by the associated Laguerre polynomials (\ref{LaguerreRec}) and are illustrated in Figure~\ref{GaudinLim}. These results are in good agreement with the direct solution of the Bethe equations, as verified with the algorithm that generated Figure~\ref{fig:BackwardsRoots}.

This limit is termed the Gaudin limit because of the similarity of the approximate BAE with the previously-presented Gaudin equations \cite{SCCG}. For $\alpha$ positive, it immediately follows from the
properties of the associated Laguerre polynomials that all $z_{j}$ are strictly positive and the rapidities lie above the line $-c_0/c_1$ for all finite values of $c_1$, as previously observed in Figure~\ref{fig:BackwardsRoots}. This expansion can also be related to the known expansion for the LL model in the limit of small $c_1$. In this case, $\alpha$ becomes large, and the roots of the associated Laguerre polynomials $L_N^{\alpha}$ can be related to the roots $y_j$ of the Hermite polynomials $H_N$ as
\beq
z^{\alpha}_{j} \approx \alpha + \sqrt{2 \alpha} y_{j} ,\qquad \textrm{for} \qquad |\alpha| \to \infty.
\eeq
Plugging this into the series expansion for $k_j$, we obtain that for small $c_1$
\beq
k_j \approx  - c_0 \sqrt{\frac{2}{L}}  y_j +\mathcal{O}(c_0^2),
\eeq
returning the known behaviour of the rapidities in the weakly-interacting Lieb-Liniger gas. For very large $c_1$, the eigenvalues become independent of $c_1$ and result in the zeros of $L_N^{-1}$. This polynomial has a single root which goes to zero as $(1+\alpha)/N$ for $\alpha \to -1$, and all other roots are non-zero for $\alpha=-1$. The exact values of these do not matter, since we see that in the limit of large $c_1$, the dominant term is given by
\beq
k_j \approx -\frac{c_0}{c_1} +\mathcal{O}(c_0^2),
\eeq
with a single rapidity behaving as
\beq
k_j \approx \frac{c_0}{c_1} (N-1)+\mathcal{O}(c_0^2).
\eeq
The existence of these two limits shows that in both cases, the series expansion was justified. Non-analytic behaviour for $c_0^2 \to 0$ is obtained, similar to the LL model, since the interaction strength is determined by $c_0^2$ and the rapidities scale as the square root of the interaction strength. From the recursion relations satisfied by the associated Laguerre polynomials, the total momentum and energy follow as
\beq
p(\{k_j\})&=&\sum_{j=1}^N k_j = 0+\mathcal{O}(c_0^2), \\
E(\{k_j\})&=&\sum_{j=1}^N k_j^2 = c_0^2 \frac{N(N-1)}{L+c_1^2}+\mathcal{O}(c_0^2).
\label{G-en}
\eeq
%%%%%%%%%%%%%%%%%%%%%%%%%%%%%%%%%%%%
\section{Bound states}
%%%%%%%%%%%%%%%%%%%%%%%%%%%%%%%%%%%%
The standard derivation of the structure of the spectrum for bound states spectrum (solutions corresponding to the complex values of rapidities) for both spin chains and continuum systems relies on the existence of a string hypothesis. This hypothesis states that in the $L\rightarrow\infty$ limit the Bethe roots are arranged into sets of self conjugate complex numbers with equal real part known as the string centre. This configuration is not conserved for finite $L$ where these strings are free to deviate from this configuration by some complex number $\delta$. For more information see \cite{Sakmann,HagermansCauxStrings,CauxCalabrese} and the references therein. The limiting cases $L\rightarrow \infty$ imposes certain constraints on the roots of the Bethe equations which are also present at large {\it finite} $L$.
Here we consider the simplest case of $N=2$ for our model and show that the string hypothesis {\it does not} hold in its original form. We conjecture a structure of bound states spectrum for general $N$.

Our derivation closely follows the one for the LL model. Namely, we start with the two-parametric Bethe equations ~(\ref{g-BA-eq})
\begin{equation}
e^{ik_j L } = \prod_{i\neq j } \frac{k_j-k_i + i ( c_0^2 + c_0 c_1 (k_j+k_i) + c_1^2 k_j k_i)}{k_j-k_i - i( c_0^2 + c_0 c_1 (k_j+k_i) + c_1^2 k_j k_i)},
\label{BAexpl}
\end{equation}
specified to $N=2$,
\beq
\begin{split}
e^{ik_{1} L}=\frac{k_1-k_2 + i ( c_0^2 + c_0 c_1 (k_1+k_2) + c_1^2 k_1 k_2)}{k_1-k_2 - i( c_0^2 + c_0 c_1 (k_1+k_2) + c_1^2 k_1 k_2)}\\
e^{ik_{2} L}=\frac{k_2-k_1 + i ( c_0^2 + c_0 c_1 (k_1+k_2) + c_1^2 k_1 k_2)}{k_2-k_1 - i( c_0^2 + c_0 c_1 (k_1+k_2) + c_1^2 k_1 k_2)}
\label{BA-2}
\end{split}.
\eeq
Introducing real and imaginary parts of $k$,
\beq
k_{j}=u_{j}+iv_{j},\qquad j=1,2
\eeq
we compare the squared modulus of both sides of the first equation in (\ref{BA-2})
\beq
e^{-2v_1 L}=\frac{A_{-}^{2}+B_{+}^{2}}{A_{+}^{2}+B_{-}^{2}}
\eeq
where
\beq
A_{\pm}&=&
u_{1}-u_{2}\pm c_{1}[c_{0}(v_{1}+v_{2})+c_{1}(u_{1}v_{2}+v_{1}u_{2})]\nonumber\\
B_{\pm}&=&v_{1}-v_{2}\pm [(c_{0}+c_{1}u_{1})(c_{0}+c_{1}u_{2})-c_{1}^{2}v_{1}v_{2}].
\label{AB}
\eeq
In the limit of $L\rightarrow\infty$ the left hand side vanishes (we postulate that $v_{1}>0$), which implies that the two conditions should be satisfied simultaneously
\beq
A_{-}=0 \qquad \mbox{and}\qquad B_{+}=0.
\eeq
Multiplying the two equations in (\ref{BA-2}) implies that $\exp[i(k_{1}+k_{2})L]=1$ which translates into
\beq
v_{1}+v_{2}=0.
\eeq
Substituting this into (\ref{AB}) gives
\beq
& &(u_{1}-u_{2})(1+c_{1}^{2}v)=0,\label{AB1}
\\
& &(c_{0}+c_{1}u_{1})(c_{0}+c_{1}u_{2})+c_{1}^{2}v^{2}+2v=0
\label{AB2}
\eeq
where $v\equiv v_{1}>0$ and so $v_{2}=-v$. The above conditions must be satisfied simultaneously.
Now, equation (\ref{AB1}) implies that either
\beq
u_{1}=u_{2}  \qquad\mbox{or}\qquad v=-\frac{1}{c_{1}^{2}}.
\label{u-v}
\eeq
In the {\bf first} case of (\ref{u-v}) equation (\ref{AB2}) can be written as an equation for a {\it circle}
\beq
 X^{2}+Y^{2}=R^{2}
\eeq
where
\beq
X&=&c_{0}+c_{1}u, \quad
Y=\frac{1}{c_{1}}+c_{1}v,\quad
 R=\frac{1}{c_{1}}.
 \eeq
This implies that the real and imaginary parts of the "string" are {\it correlated} in a special manner.

For the {\bf second} case in (\ref{u-v}) equation (\ref{AB2}) implies that $u_{1}$,$u_{2}$ and $v$ are mutually correlated,
\beq
U_{1}U_{2}=\frac{1}{c_{1}^{2}},\qquad\mbox{where}\qquad
U_{1,2}=c_{0}+c_{1}u_{1,2}.
\eeq
We note that since according to our convention $v>0$, it follows that $c_{1}$ is purely imaginary.

Interestingly, one can arrive to the same conclusion by first applying the $SL(2)$ transformation of $k_{j}$ to (\ref{BAexpl}),
\beq
k_{j,l}\rightarrow y_{j,l}=\frac{k_{j,l}}{1+\frac{c_{1}}{c_{0}}k_{j,l}}
\label{mobius}
\eeq
which transforms the right hand side of (\ref{BAexpl}) to the LL form, and then follow the same analysis as above. This is nothing but a manifestation of the geometrical property of the M\"{o}bius transformation: circles and lines (thought as circles of infinite radius) are transformed into circles and lines respectively.

By looking into the case of $N\geq3$ one can convince oneself that the requirement for numerator or denominator to vanish (which is coming from the vanishing or divergence condition respectively of the left-hand side of the Bethe equation) occurs only for a single factor in (\ref{BAexpl}) for a given $k_{j}$. This implies certain correlation between real and imaginary parts of the roots. While we cannot advance analytically anymore, we formulate a

{\bf Conjecture}: Real and imaginary parts of the roots of Bethe equations are correlated and live on domains which have proper transformation properties with respect to the action of the $SL(2)$ (like e.g. circles and lines).

This conjecture can further be supported by the Gaudin limit (see Appendix), where the BAE roots occupy a spectral curve of the Heine-Stieltjes type differential equation. Moreover it was noticed in \cite{Farey1,Farey2} that the bound state solutions in the model with $\delta$ and $\delta'$ interactions (which can be regarded as a special limit of small $c_{1}$ of the present model, such that $c_{1}^{2}$ term is neglected while the $c_{0}c_{1}$ term is kept) are situated on {\it circles} with angular positions which follow a pattern of Farey sequences in the $L=\infty$ limit. We note that the modular group $SL(2,\mathbb Z)$ is an automorphism group of the Farey sequence. It would be very interesting to study these connections further.

\begin{center}
\begin{figure}[h]
\includegraphics[width=\columnwidth]{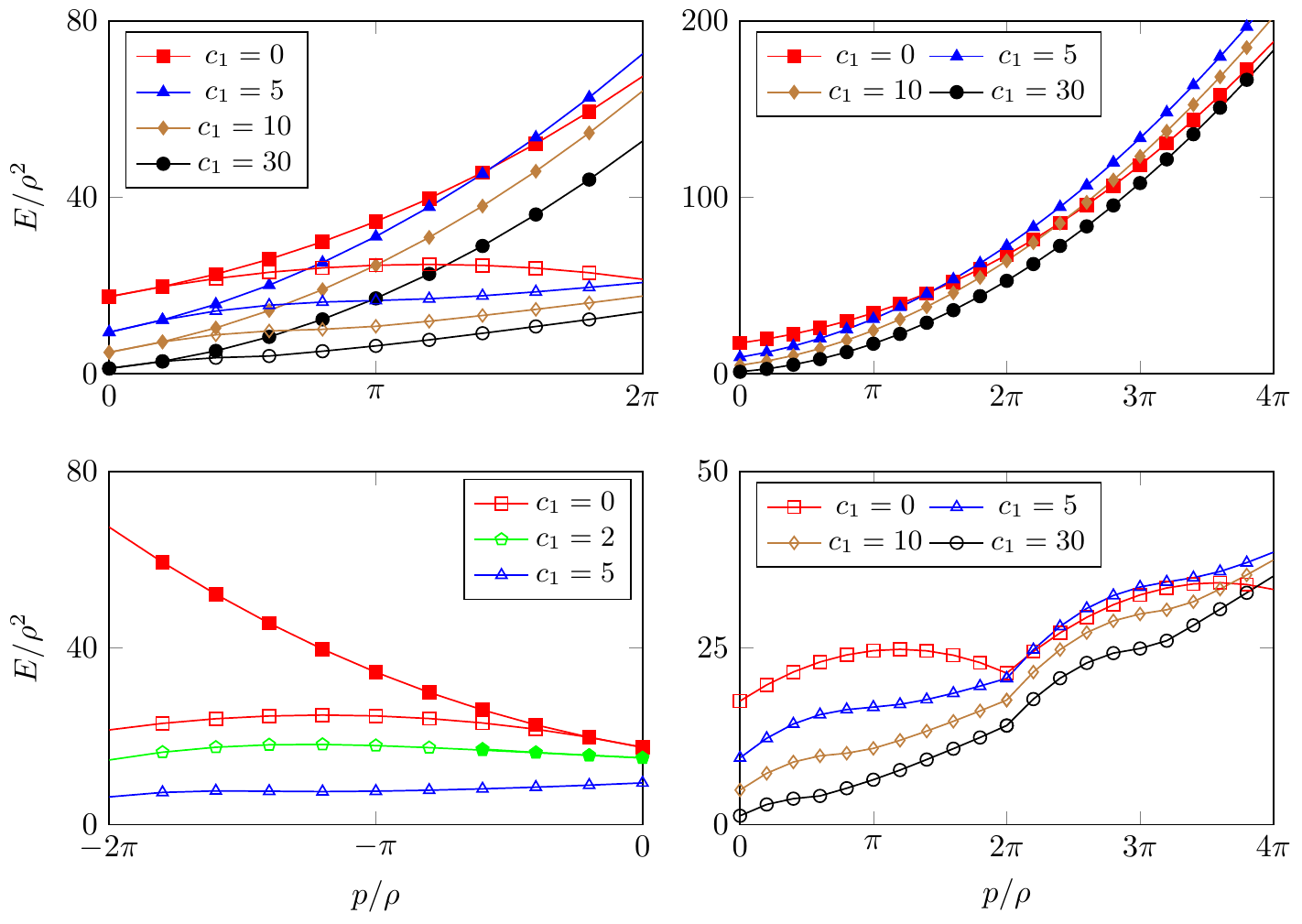}
\caption{Excitation spectrum for the ground state at $c_0 =5$ and different values of $c_1$ starting from $c_1=0$. Plots with the same value of $c_1$ are plotted with the same marker and color. \textbf{Top left}: Type-I (filled markers) and Type-II (open markers) excitation spectrum above the ground state with positive momenta. \textbf{Bottom left}: Type-I and Type-II excitations with negative momenta. At $c_1=5$ all negative momentum excited states have energies lower than those of the ground state at $p=0$ indicating that these states crossed the ground state (compare Figure~\ref{fig:Type2Spectrum}). \textbf{Top right}: Type-I excitations with large positive momenta. \textbf{Bottom right}: Type-II excitations with large positive momenta. }
\label{N10Spectrum}
\vspace{5pt}
\includegraphics[width=\columnwidth]{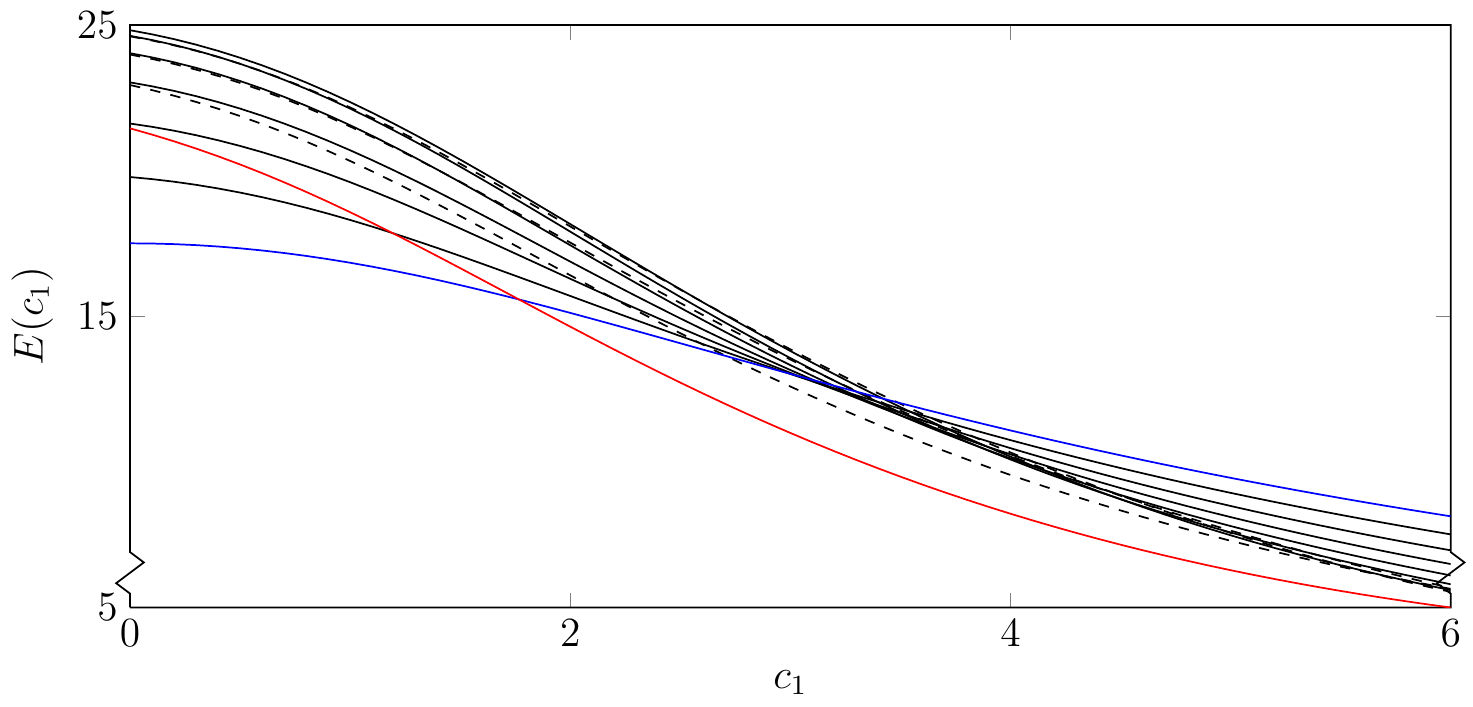}
\caption{Type-II excitation spectrum for negative-momentum excited states using $c_0=5$, $L=10$, $N=10$, starting from $c_1=0$. In blue the ground state and in red Umklapp excitation are given, crossing the ground state at $c_1 \sim 2$, where the dotted lines represent the excitations with $-2\pi<p<-\pi$.}
\label{fig:Type2Spectrum}
\end{figure}
\end{center}
%%%%%%%%%%%%%%%%%%%%%%%%%%%%%%%%%%%%%
\section{Excitations}
%%%%%%%%%%%%%%%%%%%%%%%%%%%%%%%%%%%%%
In order to discuss the excitation spectrum we distinguish between Type-I (particle) and Type-II (hole) excitations, following Lieb \cite{LL2}.  We restrict ourselves to the $c_0\gg c_1\geq0$ initial condition, in this case the negative momenta states give rise to new behavior because they approach the decoupled regime around $k_j=-c_0/c_1$. All excitations with positive momenta give similar results to the ground state because the roots never reach the line $-c_0/c_1$. For negative $c_1$ and $c_0\gg -c_1\geq0$ the root spectrum inverts around $k$ (as mentioned above) and similar behavior is observed for excitations with positive momenta whereas negative momentum excitations remain similar to the ground state.

In Figure~\ref{N10Spectrum}, the spectrum for $N=10$ is given with unit filling $N/L=1$, where again
\beq
 E(\{k_j\})= \sum_{j=1}^N k_j^2, \quad p(\{k_j\})= \sum_{j=1}^N k_j.
\eeq
Note that the choice of low $N$ is for illustrative purposes only, since large $N$ exhibits the same qualitative behavior. While the excitation spectrum remains similar for Type-I excitations for different values of $c_1$ (Figure~\ref{N10Spectrum} top right). The Type-II excitations lose their characteristic quadratic behavior and tend to $E \sim p$ instead of $E\sim |p|$ as expected for fermions. Comparing the bottom panels of Figure~\ref{N10Spectrum} the energies of the Type-II excitations for larger $c_1$ becomes strictly increasing with respect to the momenta, the linear behavior becomes more obvious for increasing $c_1$. This behavior is confirmed by the evaluation of the energy of negative momenta Type-II excitations as a function of $c_1$ in Figure~\ref{fig:Type2Spectrum}. A full inversion of the spectrum occurs at $c_1\sim3.5$, which means that the spectrum reorders and becomes strictly increasing as expected.

Unfortunately, we were unable to continue our analysis of the negative momenta states to increasingly negative momenta or higher $c_1$ because these have the same numerical instabilities as the $c_1 \gg c_0$ initial condition states near $c_1\rightarrow 0$, where some of the rapidities hit the line $k=-c_0/c_1$.

%\begin{center}
%\begin{figure}
%\includegraphics[width=4cm]{fun_N100_c5}
%\includegraphics[width=4cm]{fun_N100_c05}\\
%\includegraphics[width=4cm]{deriv_N100_c5}
%\includegraphics[width=4cm]{deriv_N100_c05}\\
%\includegraphics[width=4cm]{deriv2}
%\includegraphics[width=4cm]{deriv2_N100_c05}\\
% \includegraphics[width=4cm]{deriv3}
% \includegraphics[width=4cm]{deriv3_N100_c05}\\
%\caption{ The energy of the ground state and its first three derivatives as a function of $c_{1}$ at $c_{0}=5$ (left panel) and $c_{0}=0.5$ (right panel)\label{fig:en}}
%\end{figure}
%\end{center}

%%%%%%%%%%%%%%%%%%%%%%
\section{Conclusion}
%%%%%%%%%%%%%%%%%%%%
Based on a previously-presented general solution of the rational Yang-Baxter equation~\cite{SCCG},  we {\it initiated} the studies of a generalized model of an interacting one-dimensional gas. Our first guess was that these particles can, {\it presumably}, be interpreted as anyons in the spirit of Refs. \cite{BGH,GBK,Kundu}, since the different limits of this model return either interacting bosons with $\delta$ interaction or CS model with $\delta''$ interaction. However if our conjectured form of the Hamiltonian (\ref{HF}) is correct, our model may be seen as the one containing density-current (mixed $c_{0}c_{1}$ term) and current-current interactions ($c_{1}^{2}$ term), in addition to the density-density coupling (traditional $c_{0}^{2}$ term). The $SL(2)$ acts then as a duality between density and current. This interpretation partially explains the title of this paper.

The interaction structures follow from study of the contact interactions \cite{Seba,Albeverio,rev-VS,Bajnok} which can be related to different subgroups of $SL(2)$. The standard ($\delta$) interaction corresponds to the upper parabolic block of the transformation matrix (Borel subalgebra) while the lower parabolic group, related by the $\mathcal S$-duality with the former, is given by the lower parabolic block and corresponds to the $\delta''$ interaction. The diagonal subalgebra corresponds to either the elliptic or hyperbolic block and is connected with $\delta$ interaction.

The Bethe ansatz equations are obtained assuming that there exists {\it a basis} where the permutation operator acts as an identity, and are solved both numerically and by connecting the rapidities to roots of orthogonal polynomials in the weakly interacting limit. Also the excitation spectrum is presented.
A particular feature of the presented model is the presence of non-interacting modes at specific values of the interaction constants, which prevents the ground states of the corresponding fermionic and bosonic limiting models from being smoothly connected. These modes are connected by the modular $SL(2,\mathbb{Z})$ subgroup of $SL(2)$.

We further conjecture that
the string hypothesis should perhaps be generalized to more general domains of the rapidities complex plane. We believe that future studies of the presented model will encounter new surprises and mathematical richness.

{\it Acknowledgements.}
This work is part of the Delta-ITP consortium, a program of the Netherlands Organization for Scientific Research (NWO) that is funded by the Dutch Ministry of Education, Culture and Science (OCW). P.W.C. acknowledges support from a Ph.D. fellowship and a travel grant for a long stay abroad at the University of Amsterdam from the Research Foundation Flanders (FWO Vlaanderen). J.-S. C. acknowledges support from  the European Research Council under ERC Advanced grant 743032 DYNAMINT.

\section{Appendix}

\subsection{Derivation of the integral equation \label{LL-int-eq}}

Here we outline a derivation of the integral equation in the continuum limit $L\rightarrow\infty$, $N\rightarrow\infty$, $N/L=\rho_{0}=const$, though we believe that it has limited application because of singularities contained in the kernel. These singularities can be extracted to the driving term by use of the transformation (\ref{mobius}) which cleans up the kernel but gives the same singularities.

We start with the equation (\ref{g-BA-eq}) in its explicit form
% \beq
% e^{ik_{j}L}=\prod_{l\neq j}^{N}\frac{k_{j}-k_{l}+ic_{0}\left(1+\frac{c_{1}}{c_{0}}(k_{j}+k_{l})+\frac{c_{1}^{2}}{c_{0}^{2}}k_{j}k_{l}\right)}{k_{j}-k_{l}-ic_{0}\left(1+\frac{c_{1}}{c_{0}}(k_{j}+k_{l})+\frac{c_{1}^{2}}{c_{0}^{2}}k_{j}k_{l}\right)}.
% \label{BA-eq-expl}
% \eeq
\beq
e^{ik_{j}L}=\prod_{l\neq j}^{N}\frac{k_{j}-k_{l}+i(c_0+c_{1}k_{j})(c_0+c_1k_l)}{k_{j}-k_{l}-i(c_0+c_{1}k_{j})(c_0+c_1k_l)}.
\label{BA-eq-expl}
\eeq
The usual strategy of analyzing these coupled transcendental equations is to consider a logarithm of its both sides
\beq
L k_{j}+\sum_{l=1}^{N}\theta(k_{j},k_{l})=2\pi \left(n_{j}-\frac{N+1}{2}\right)
\eeq
where $\{ n_{j}\}$, $j=1,\ldots, N$ is a set of integers, and
% \beq
% \theta(k_{j},k_{l})=i\ln \left(\frac{ic_{0}f(k_{j},k_{l})+(k_{j}-k_{l})}{ic_{0}f(k_{j},k_{l})-(k_{j}-k_{l})}\right)\nonumber\\
% \eeq
\beq
\theta(k_{j},k_{l})=i\ln \left(\frac{if(k_{j},k_{l})+(k_{j}-k_{l})}{if(k_{j},k_{l})-(k_{j}-k_{l})}\right)\nonumber\\ \label{Kernel}
\eeq
is a scattering phase shift. Here
% \beq
% f(k_{j},k_{l})=\left(1+\frac{c_{1}}{c_{0}}(k_{j}+k_{l})+\frac{c_{1}^{2}}{c_{0}^{2}}k_{j}k_{l}\right).
% \eeq
\beq
f(k_{j},k_{l})=c_0^2+c_{1}c_{0}(k_{j}+k_{l})+c_{1}^{2}k_{j}k_{l}.
\eeq
The kernel (\ref{Kernel}) is an antisymmetric function and therefore
\beq
\sum_{j,l=1}^{N}\theta(k_{j},k_{l})=0,
\eeq
which implies that
\beq
L\sum_{j=1}^{N}k_{j}=2\pi\sum_{j=1}^{N}\left(n_{j}-\frac{N+1}{2}\right)
\eeq
and we can interpret $\sum_{j=1}^{N}k_{j}$ as a total momentum.

One can bring the phase shift to the Lieb-Liniger form by applying a particular $SL(2)$ transformation
\beq
k_{j,l}\rightarrow y_{j,l}=\frac{k_{j,l}}{1+\frac{c_{1}}{c_{0}}k_{j,l}}.
\eeq
By considering the extended complex plane of $k_{j,l}$ augmented by the point at infinity, we can formally allow singularities in the kernel $\theta(k_{j},k_{l})$. In terms of the $y_{j}$-variables the Bethe Ansatz equation gets the following form
\beq
L\left(\frac{y_{j}}{1-\frac{c_{1}}{c_{0}}y_{j}}\right)+\sum_{k=1}^{N}\theta_{LL}(y_{j}-y_{l})=2\pi\left(j-\frac{N+1}{2}\right)\nonumber\\\label{LBAEy}
\eeq
where
\beq
\theta_{LL}(k)=i\ln\left(\frac{ic_{0}^2+k}{ic_{0}^2-k}\right).
\eeq
Introducing the root density as
\beq
\rho(y_{j})&=&\frac{1}{L(y_{j+1}-y_{j})},\\
\sum_{j=1}^{N}&=&L\sum_{j=1}^{N}\rho(y_{j})(y_{j+1}-y_{j})\\
\sum_{j=1}^{N}f(y_{j})&\rightarrow &L\int_{A}^{B}dy\rho(y)f(y)
\eeq
for arbitrary function $f(y)$, we subtract equation (\ref{LBAEy}) for $y_{j+1}$ from the same equation for $y_{j}$ and obtain
\beq
L\frac{c_{0}}{c_{1}}\left(\frac{1}{1-\frac{c_{1}}{c_{0}}y_{j+1}}-\frac{1}{1-\frac{c_{1}}{c_{0}}y_{j}}\right)\\
+(y_{j+1}-y_{j})\sum_{l=1}^{N}\theta'(y_{j}-y_{l})=2\pi
\eeq
as $N,L\rightarrow\infty$. Here $\theta_{LL}'(y_{j}-y_{m})(y_{j+1}-y_{j})=\theta_{LL}(y_{j+1}-y_{m})-\theta_{LL}(y_{j}-y_{m})$ and we also {\it assume } that the boundaries of integration $A=-B=-\Lambda$.
Finally, expanding around small $\delta y =(y_{j+1}-y_j)=\frac{1}{\rho(y_j)L}$ and {\it assuming} that terms of $\mathcal{O}(\delta y^2)$ can be neglected
\beq
\left(\frac{1}{1-\frac{c_{1}}{c_{0}}y_{j+1}}-\frac{1}{1-\frac{c_{1}}{c_{0}}y_{j}}\right)\approx \frac{c_{1}}{c_{0}L}\frac{1}{\rho(y)(1-\frac{c_{1}}{c_{0}}y)^{2}}\nonumber\\
\eeq
% (that is neglecting terms like $1/\rho(y)$ and $y/\rho(y)$ in the denominator)
we obtain
\beq
\rho(y)-\frac{1}{2\pi}\int_{-\Lambda}^{\Lambda} dx \rho(x)K(x,y)=\frac{1}{2\pi}\frac{1}{(1-\frac{c_{1}}{c_{0}}y)^{2}}
\label{LL-gen-eq}
\eeq
where
% \beq
% K(x,y)=\frac{2c_{0}}{c_{0}^{2}+(x-y)^{2}}
% \eeq
\beq
K(x,y)=\frac{2c_{0}^2}{c_{0}^{4}+(x-y)^{2}}
\eeq
and we have to supplement (\ref{LL-gen-eq}) by the normalization condition
\beq
\int_{-\Lambda}^{\Lambda}\rho(y)dy=D=\frac{N}{L}.
\eeq
The final step, which is very convenient for the numerical implementation is to perform the following change of variables,
\beq
c_{0}=\Lambda^{\frac{1}{2}}\alpha,\quad c_0 c_{1}\equiv \beta^2,\quad y=\Lambda k, \quad \rho(\Lambda y)=g(k)
\eeq
in terms of which we have the following system
\beq
g(k)-\frac{\alpha^2}{\pi}\int_{-1}^{1}\frac{g(p)dp}{\alpha^{4}+(k-p)^{2}}&=&\frac{1}{2\pi}\frac{1}{(1-\frac{\beta^2}{\alpha^2}k)^{2}} \\
\gamma\int_{-1}^{1}dk g(k)&=&\alpha^2
\eeq
where $\gamma=c_{0}^2/D$.

Alternatively, one can perform the same analysis without making use of the $SL(2)$ transform while dealing directly with the modified kernel (with respect to the LL kernel)
\begin{equation}
\tilde{K}(k,q)=\frac{(\alpha^2+\beta^2 k)(\alpha^2+\beta^2 q)-\beta^2(k-q)(\alpha^2+\beta^2 q)}{(\alpha^2+\beta^2 k)^{2}(\alpha^2+\beta^2 q)^{2}+\alpha^{4}(k-q)^{2}}.
\end{equation}
We end up with the following equation then
\begin{equation}
g(k)-\frac{\alpha^2}{\pi}\int_{-1}^{1}dq g(q) \tilde{K}(k,q)=\frac{1}{2\pi}\\
\end{equation}
supplemented by $\gamma\int_{-1}^{1}dk g(k)=\alpha^2$

The assumption that we can perform a Taylor expansion for arriving to the integral equation is crucial here. This assumption is however only justified if the $\frac{c_0}{c_1}>|\Lambda|$ or $\frac{\beta}{\alpha}<1$ which in general is not true for all parameters $c_{0,1}$. At these same points the modified kernel becomes singular which is also reflected in a singular behavior of the finite density, Fig.~(\ref{fig:Density}). An attempt was made to solve these equations numerically, for the case where $\frac{c_0}{c_1}>|\Lambda|$ the numerics correspond to the finite density behavior, however we were unable to continue our studies outside this regime.

%\ves{For small $c_0$ I was able to reproduce the results of Pieter shown in the following figures up to the divergence in imaginary part. Unfortunately Mathematica throws only errors (precision \& singular Jacobian) for larger $c_1$ so I was unable to obtain anything new. The four particle case is for the solution using the regular Bethe equations, here we see non-zero real part before the complex part diverges surprisingly. Pieter \& Vladimir, maybe we need to talk about this stuff still...}\qpwc{Maybe we can leave this out of the paper, it is getting crowded enough as is...}

\subsection{Gaudin limit}
Starting from the Bethe ansatz equations (\ref{g-BA-eq})
\beq
e^{i k_j L} = \prod_{l \neq j} \frac{k_j-k_l+i (c_0+c_1 k_j)(c_0+c_1 k_l)}{k_j-k_l -i (c_0+c_1k_j)(c_0+c_1 k_l)},\label{BAEGaudin}
\eeq
a series expansion for the rapidities can be obtained in orders of $c_0$ if $c_0 \ll 1$. We propose an expansion similar to the weakly-interacting limit of the Lieb-Liniger gas with
\beq
k_j = c_0 \delta_j+\mathcal{O}(c_0^2),
\eeq
in which $j=1 \dots N$ and $\delta_j$ is finite and independent of $c_0$. Up to zeroth order in $c_0$, the BAE becomes $1=1$, which is trivially satisfied. The first-order correction on the equations (\ref{BAEGaudin}) is then given by
\beq
\delta_j L = 2  \sum_{l \neq j}^N \frac{(1+c_1\delta_j)(1+c_1 \delta_l)}{\delta_j-\delta_l}.\label{FOC}
\eeq
In taking this series expansion, we have made the assumption that $c_1 \delta_j$ remains bounded for all values of $c_1$. This is expected for small $c_1$ and can be verified afterwards for large $c_1$. The equation (\ref{FOC}) can again be linked to the roots of orthogonal polynomials. Performing the substitution $\delta_j = -\frac{1}{c_1} + \frac{L}{c_1^3 z_j}$ and multiplying the equation with $-c_1/L$ results in
\beq
1-\frac{L/c_1^2}{z_j} = 2 \sum_{l \neq j}^N \frac{1}{z_j-z_l}.
\eeq
The associated Laguerre polynomials $L^{\alpha}_N(z)$ satisfy the differential equation
\beq
z P''(z)+(1+\alpha-z)P'(z)+N P(z) = 0,
\eeq
where from the Heine-Stieltjes connection \cite{szego_orthogonal_1975} the roots $z_j$ are coupled through
\beq
1-\frac{1+\alpha}{z_j} = 2\sum_{l \neq j}^N\frac{1}{z_j-z_l}.
\eeq
Taking $1+\alpha = L/c_1^2$, this returns the proposed series expansion as
\beq
k_j = -\frac{c_0}{c_1}+\frac{c_0}{c_1^3}\frac{L}{z_j} + \mathcal{O}(c_0^2),
\eeq
with $z_j$ the $j$-th root of the associated Laguerre polynomial $L_N^{\alpha}(z)$ with $1+\alpha = L/c_1^2$. This is known as a Laguerre function for non-integer values of $\alpha$, satisfying the following recursion relation
\beq
N L_N^{\alpha}(x) &=& (2N-1+\alpha-x)L_{N-1}^{\alpha}(x) \nonumber \\
&&- (N-1+\alpha)L_{N-2}^{\alpha}(x), \label{LaguerreRec}
\eeq
from which the roots can be determined for arbitrary values of $\alpha$. These also satisfy
\beq
L_N^{\alpha}(0) = \genfrac(){0pt}{0}{N+\alpha}{N},
\eeq
and derivatives can be found as
\beq
\frac{\mathrm{d}^k}{\mathrm{d}x^k}L_N^{\alpha}(x) = (-1)^k L^{\alpha+k}_{N-k}(x),
\eeq
for $k<N$. The roots of the polynomial $P(z)=L_N^{\alpha}(z)$ can then be easily shown to satisfy
\beq
\sum_{j=1}^N \frac{1}{z_j} &=& -\frac{P'(0)}{P(0)}  = \frac{N}{1+\alpha}, \\
\sum_{j=1}^N \frac{1}{z_j^2} &=& \left(\frac{P'(0)}{P(0)} \right)^2-  \frac{P''(0)}{P(0)} = \frac{N(N+1+\alpha)}{(1+\alpha)^2(2+\alpha)} \nonumber.
\eeq
Using this in the expansion for the rapidities then returns the results presented in the main text.

\begin{center}
\begin{figure}
\includegraphics[width=0.47\textwidth]{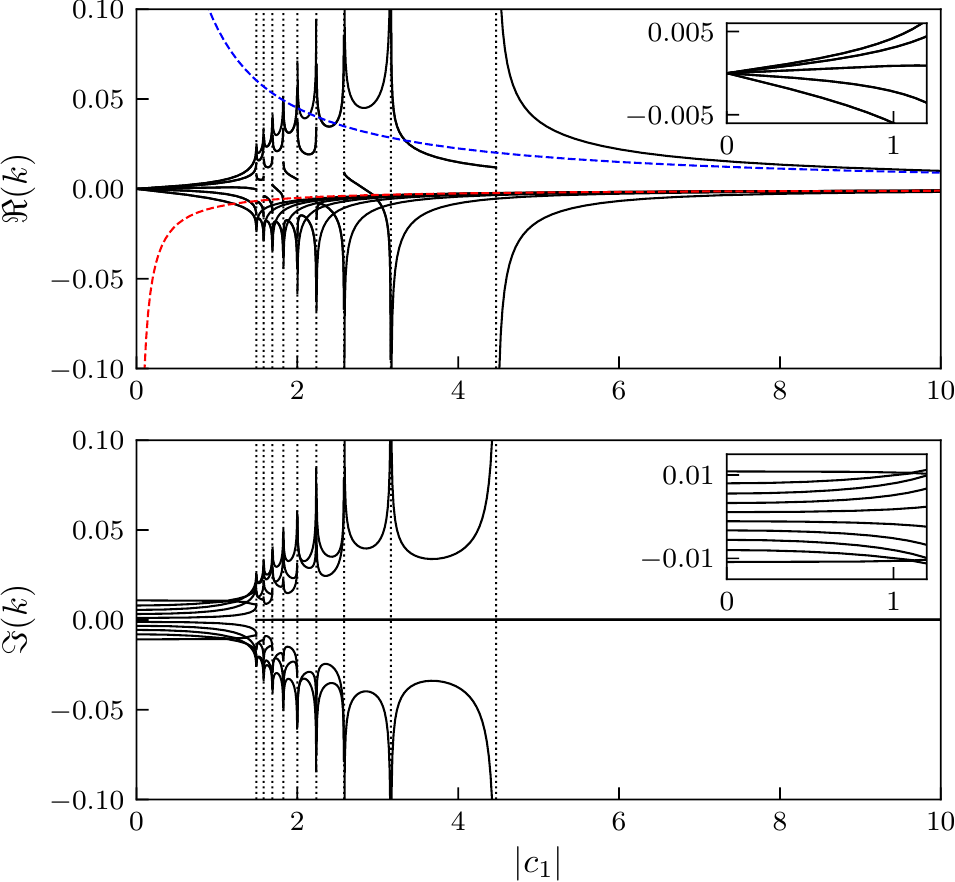}
\caption{Series approximation to the rapidities for purely imaginary parameters with $|c_0| = 0.01$, $L=20$ and $n=10$. The red dashed line denotes $-c_0/c_1$, the blue dashed line denotes $(n-1)c_0/c_1$, and the vertical dashed lines denote $|c_1|^2 = L/m,m =1 \dots n$\label{LLneg}. Insets show a zoom of the main plots at small values of $|c_1|$ for comparison with the attractive LL model.
}\end{figure}
\end{center}

Interesting things happens if we plot the rapidities $k$ for both $c_0$ and $c_1$ purely imaginary, as done in Fig.~(\ref{LLneg}). Now the rapidities start as purely imaginary for $c_1=0$ as expected for the attractive LL case \cite{HagermansCauxStrings,CauxCalabrese}, but obtain a real part for non-zero $c_1$. Remarkably, the approximation breaks down at values of $c_1^2 = -L/m, m =1 \dots n$. At these points, the Laguerre polynomials have zero as root, and the first-order contribution diverges. It should be checked numerically what happens at these points if we solve the full Bethe equations rather than the first-order approximation. This also allows for a crossing of the line $-c_0/c_1$ if we consider the large $c_1$ limit. Again, at both small and large $c_1$ the approximation holds, but at intermediate $c_1$ the approximation breaks down at exactly $c_1^2 = -L/m, m =1 \dots n$, where the numerical error on the Bethe equations becomes large. However, if we plot the energy, these divergences cancel exactly except for $c_1^2 = -L$, where the energy first dives towards minus infinity for $|c_1|^2<L$ and returns from plus infinity for $|c_1|^2>L$. This is consistent with the resulting expression for the
total energy (\ref{G-en}).

\begin{thebibliography}{99.}

\bibitem{LL1}
E. H. Lieb and W. Liniger, Exact Analysis of an Interacting Bose Gas. I. The General Solution and the Ground State,
\href{http://dx.doi.org/10.1103/PhysRev.130.1605}{Phys. Rev. {\bf 130}, 1605 (1963).}

\bibitem{LL2}
E. H. Lieb, Exact Analysis of an Interacting Bose Gas. II. The Excitation Spectrum,
\href{https://doi.org/10.1103/PhysRev.130.1616}{Phys. Rev. {\bf 130}, 1616 (1963).}

\bibitem{Lai1}
Y. Lai and H. A. Haus, Quantum theory of solitons in optical fibers. I. Time-dependent Hartree approximation, \href{http://dx.doi.org/10.1103/PhysRevA.40.844}{
Phys. Rev. A {\bf 40}, 844 (1989).}

\bibitem{Lai2}
Y. Lai and H. A. Haus, Quantum theory of solitons in optical fibers. II. Exact solution, \href{http://dx.doi.org/10.1103/PhysRevA.40.854}{Phys. Rev. A  {\bf 40}, 854 (1989).}


\bibitem{Yudson}
V. I. Yudson, Dynamics of Integrable Quantum Systems, \href{http://jetp.ac.ru/cgi-bin/dn/e_061_05_1043.pdf}{Zh. Eksp. Teor. Fiz. {\bf 88} (1984) 1757 (1985). [Soy. Phys. JETP {\bf 61} 1043 (1985).}

\bibitem{Imamoglu}
I. Carusotto, D. Gerace, H. E. Tureci, S. De Liberato, C. Ciuti, and A. Imamoglu, Fermionized Photons in an Array of Driven Dissipative Nonlinear Cavities, \href{http://dx.doi.org/10.1103/PhysRevLett.103.033601}{Phys. Rev. Lett. {\bf 103}, 033601 (2009).}

\bibitem{Chang}
D. Chang, V. Gritsev, G. Morigi, M. Lukin, E. Demler, Crystallization of strongly interacting photons in a nonlinear optical fibre, \href{http://dx.doi.org/10.1038/nphys1074}{Nature Phys. {\bf 4}, 884 (2008).}

\bibitem{Kardar}
M. Kardar, Replica Bethe Ansatz studies of two-dimensional interfaces with quenched random impurities,
\href{http://dx.doi.org/10.1016/0550-3213(87)90203-3}{Nucl. Phys. B {\bf 290}, 582 (1987).}

\bibitem{KPZ}
M. Kardar, G. Parisi, and Y.-C. Zhang, Dynamic Scaling of Growing Interfaces, \href{http://dx.doi.org/10.1103/PhysRevLett.56.889}{
Phys. Rev. Lett. {\bf 56}, 889 (1986).}

\bibitem{CD}
P. Calabrese and P. Le Doussal, Exact Solution for the Kardar-Parisi-Zhang Equation with Flat Initial Conditions, \href{http://dx.doi.org/10.1103/PhysRevLett.106.250603}{Phys. Rev. Lett. {\bf 106}, 250603 (2011).}

\bibitem{Dotsenko}
V. Dotsenko, Bethe Ansatz derivation of the Tracy-Widom distribution for one-dimensional directed polymers,
\href{http://dx.doi.org/10.1209/0295-5075/90/20003}{EPL {\bf 90}, 20003  (2010).}


\bibitem{TW}
C. A. Tracy and H. Widom, The Bose Gas and Asymmetric Simple Exclusion Process on the Half-Line, 	\href{https://dx.doi.org/10.1007/s10955-012-0686-4}{J. Stat. Phys. (2013) {\bf 150}, 1 (2013).}


\bibitem{Olshanii_98}
M. Olshanii, Atomic Scattering in the Presence of an External Confinement and a Gas of Impenetrable Bosons,
\href{https://link.aps.org/doi/10.1103/PhysRevLett.81.938}{Phys. Rev. Lett. {\bf 81}, 938 (1998).}

\bibitem{KWW1}
T. Kinoshita, T. Wenger, and D. S. Weiss, Observation of a One-Dimensional Tonks-Girardeau Gas, \href{http://dx.doi.org/10.1126/science.1100700}{Science {\bf 305}, 1125 (2004). }

\bibitem{KWW2}
T. Kinoshita, T. Wenger, and D. S. Weiss, Local Pair Correlations in One-Dimensional Bose Gases, \href{http://dx.doi.org/10.1103/PhysRevLett.95.190406}{Phys. Rev. Lett. {\bf 95}, 190406 (2005).}

\bibitem{Bloch}
B. Paredes, A. Widera, V. Murg, O. Mandel, S. F\"olling, I. Cirac, G. V. Shlyapnikov, T. W. H\"ansch, and I. Bloch, Tonks-Girardeau gas of ultracold atoms in an optical lattice,
\href{http://dx.doi.org/10.1038/nature02530}{Nature {\bf 429}, 277 (2004).}

\bibitem{Tolra_04}
B. Laburthe Tolra, K. M. O'Hara, J. H. Huckans, W. D. Phillips, S. L. Rolston, and J. V. Porto,
Observation of Reduced Three-Body Recombination in a Correlated 1D Degenerate Bose Gas,
\href{https://link.aps.org/doi/10.1103/PhysRevLett.92.190401}{Phys. Rev. Lett. {\bf 92}, 190401 (2004).}

\bibitem{Haller_11}
E. Haller, M. Rabie, M. J. Mark, J. G. Danzl, R. Hart, K. Lauber, G. Pupillo, and H.-C. N\"agerl,
Three-Body Correlation Functions and Recombination Rates for Bosons in Three Dimensions and One Dimension,
\href{https://link.aps.org/doi/10.1103/PhysRevLett.107.230404}{Phys. Rev. Lett. {\bf 107}, 230404 (2011).}

\bibitem{Kinoshita_06}
T. Kinoshita, T. Wenger, and D. S. Weiss,
A quantum Newton's cradle,
\href{http://dx.doi.org/10.1038/nature04693}{Nature {\bf 440}, 900 (2006).}

\bibitem{Haller_09}
E. Haller, M. Gustavsson, M. J. Mark, J. G. Danzl, R. Hart, G. Pupillo, and H.-C. N\"agerl,
Realization of an Excited, Strongly Correlated Quantum Gas Phase,
\href{http://dx.doi.org/10.1126/science.1175850}{Science {\bf 325}, 1224 (2009).}

\bibitem{Meinert_17}
F. Meinert, M. Knap, E. Kirilov, K. Jag-Lauber, M. B. Zvonarev, E. Demler, and H.-C. N\"agerl,
Bloch oscillations in the absence of a lattice,
\href{http://dx.doi.org/10.1126/science.aah6616}{Science {\bf 356}, 945 (2017).}

\bibitem{rev}
M. A. Cazalilla, R. Citro, T. Giamarchi, E. Orignac, and M. Rigol, One dimensional bosons: From condensed matter systems to ultracold gases, \href{http://dx.doi.org/10.1103/RevModPhys.83.1405}{Rev. Mod. Phys. {\bf 83}, 1405 (2011).}


\bibitem{Gaudin1}
M.~Gaudin, Un systeme a une dimension de fermions en interaction, \href{https://doi.org/10.1016/0375-9601(67)90193-4}{Phys. Lett. A, {\bf 24}, 55 (1967).}

\bibitem{Yang1}
C.N. Yang, Some exact results for the many-body problem in one dimension with repulsive delta-function interaction, \href{http://dx.doi.org/10.1103/PhysRevLett.19.1312}{Phys. Rev. Lett. {\bf 19}, 1312 (1967).}

\bibitem{Yang2}
C.N. Yang, S matrix for the one-dimensional N-body problem with repulsive-function interaction, \href{http://dx.doi.org/10.1103/PhysRev.168.1920}{Phys. Rev. {\bf 168}, 1920 (1968).}


\bibitem{Sutherland1}
B. Sutherland, Further Results for the Many-Body Problem in One Dimension, \href{https://doi.org/10.1103/PhysRevLett.20.98}{Phys. Rev. Lett. {\bf 20}, 98 (1968)}.

\bibitem{SCCG}
E. Stouten, P. W. Claeys, J-S. Caux and V. Gritsev, Integrability and duality in spin chains, \href{https://arxiv.org/abs/1712.09375}{arXiv:1712.09375 (2018)}.


\bibitem{CS-model}
T. Cheon and T. Shigehara, Fermion-Boson Duality of One-Dimensional Quantum Particles with Generalized Contact Interactions, \href{http://dx.doi.org/10.1103/PhysRevLett.82.2536}{Phys. Rev. Lett. {\bf 82}, 2536 (1999).}

\bibitem{CS-potential}
T. Cheon and T. Shigehara, Realizing discontinuous wave functions with renormalized short-range potentials, \href{http://dx.doi.org/10.1016/S0375-9601(98)00188-1}{Phys. Lett. A {\bf 243}, 111 (1998).}

\bibitem{BGH}
M. T. Batchelor, X.-W. Guan, and J.-S. He, The Bethe Ansatz for 1d interacting anyons,
\href{http://stacks.iop.org/1742-5468/2007/i=03/a=P03007}{J. Stat. Mech.: Th. Exp., 2007(03):P03007, 2007}.

\bibitem{GBK}
X.-W. Guan M.T. Batchelor and A. Kundu, One-dimensional anyons with competing-function and derivative-function potentials, \href{http://stacks.iop.org/1751-8121/41/i=35/a=352002}{J. Phys. A: Math. Theor. {\bf 41}, 352002 (2008)}.

\bibitem{Kundu}
A. Kundu, Exact Solution of Double $\delta$-Function Bose Gas through an Interacting Anyon Gas, \href{http://dx.doi.org/10.1103/PhysRevLett.83.1275}{Phys. Rev. Lett. {\bf 83}, 1275 (1999).}



%\bibitem{Feiguin}
%A. Feiguin, S. Trebst, A. W. W. Ludwig, M. Troyer, A. Kitaev, Z. Wang, M. H. Freedman, Interacting anyons in topological quantum liquids: The golden chain, \href{http://dx.doi.org/10.1103/PhysRevLett.98.160409}{Phys. Rev. Lett. {\bf 98}, 160409 (2007).}



%self adjoiunt

\bibitem{Seba}
P. $\check{\mbox{S}}$eba, The generalized point interaction in one dimension, \href{http://dx.doi.org/10.1007/BF01597402}{Czech. J. Phys. B, {\bf 36}, 667 (1986).}

\bibitem{Albeverio}
S. Albeverio, F. Gesztesy, R. Hoegh-Krohn and H. Holden, {\it Solvable Models in Quantum Mechanics} (Springer, Heidelberg, 1988).


\bibitem{rev-VS}
S. De Vincenzo and C. S\'{a}nchez, Point interactions: boundary conditions or potentials with the Dirac delta function,
\href{http://dx.doi.org/10.1139/P10-060}{Can. J. Phys. {\bf 88}, 809 (2010).}

\bibitem{Bajnok}
L. $\check{\mbox{S}}$amaj and Z. Bajnok, {\it Introduction to the Statistical Physics of Integrable Many-body Systems} (Cambridge UP, 2013).








%Gaudin-Richardson


%\bibitem{Links1}
%J. Links, \href{https://arxiv.org/pdf/1206.2988.pdf}{arXiv:1206.2988}

%%Library Pieter
%\bibitem{dukelsky_colloquium:_2004}
%J.~Dukelsky, S.~Pittel, and G.~Sierra,
%\newblock {C}olloquium: {Exactly} solvable {Richardson}-{Gaudin} models for many-body quantum systems,
%\newblock \href{10.1103/RevModPhys.76.643}{Rev. Mod. Phys. {\bf 76}, 643 (2004).}


%\bibitem{amico_integrable_2001}
%L.~Amico, A.~Di~Lorenzo, and A.~Osterloh,
%\newblock {I}ntegrable {Model} for {Interacting} {Electrons} in {Metallic} {Grains},
%\newblock \href{10.1103/PhysRevLett.86.5759}{ Phys. Rev. Lett. {\bf 86}, 5759 (2001).}


%\bibitem{ortiz_exactly-solvable_2005}
%G.~Ortiz, R.~Somma, J.~Dukelsky, and S.~Rombouts,
%\newblock{E}xactly-solvable models derived from a generalized {Gaudin} algebra.,
%\newblock \href{10.1016/j.nuclphysb.2004.11.008}{Nucl. Phys. B {\bf 707}, 421 (2005)}


\bibitem{gaudin_bethe_2014}
M.~Gaudin and J.-S.~ Caux,\newblock {\em {T}he {Bethe} {Wavefunction}},
\newblock Cambridge University Press, Cambridge, 2014.


\bibitem{szego_orthogonal_1975}
G.~Szeg\"o,\newblock {\em {Orthogonal} {Polynomials}, 4th edn.},
\newblock American Mathematical Society, Providence, RI, 1975.


\bibitem{Sakmann}
K. Sakmann, A. I. Streltsov, O. E. Alon, and L. S. Cederbaum, Exact ground state of finite Bose-Einstein condensates on a ring, \href{https://doi.org/10.1103/PhysRevA.72.033613}{Phys. Rev. A {\bf 72}, 033613 (2005).}


\bibitem{HagermansCauxStrings}
R.~Hagemans and J.-S.~Caux, \newblock{D}eformed strings in the Heisenberg model, \newblock \href{http://iopscience.iop.org/article/10.1088/1751-8113/40/49/001}{{J. Phys. A: Math. Theor.},{\bf 40}, 14605-14647 (2007)}.

\bibitem{CauxCalabrese}
P.~Calabrese and J.-S.~Caux, \newblock{Dynamics of the attractive 1D Bose gas: analytical treatment from integrability,} \href{http://stacks.iop.org/1742-5468/2007/i=08/a=P08032}{{J. Stat. Mech}, {\bf 08} P08032 (2007)}.

\bibitem{Farey1}
B. Basu-Mallick, T. Bhattacharyya, and D. Sen, Multi-band structure of a coupling constant for quantum bound
states of a generalized nonlinear Schr\"{o}dinger model, \href{http://dx.doi.org/10.1016/j.physleta.2005.05.021}{Phys. Lett. A {\bf 341}, 371 (2005).}

\bibitem{Farey2}
B. Basu-Mallick, T. Bhattacharyya, and D. Sen, Clusters of bound particles in a quantum integrable many-body
system and number theory, \href{https://doi.org/10.1088/1742-6596/563/1/012003}{J. Phys.: Conf. Ser. {\bf 563}, 012003 (2014).}


\bibitem{foot1}
Some preliminary steps where taken to show that the Hamiltonian (\ref{HF}) indeed results in the desired Bethe equations (\ref{g-BA-eq}) with (\ref{S-matrix}) and (\ref{F-new}) in a similar spirit as \cite{BGH}. This process for the potential (\ref{V}) however has proven to be more subtle as in \cite{BGH}. Because of the highly singular nature of the potential one has to deal with generalized functions (distributions) with great care, and therefore define a corresponding Hilbert space and matrix elements which could depend on regularization procedure. In order not to overload the current paper this analysis will be continued in future work, in this sense the Hamiltonian is conjecture.

\bibitem{foot2}
The statement that $\Pi_{ij}=1$ is a different way of saying that one is dealing with indistinguishable particles. In contrast, if one would consider ''colored'' particles a more general representation of the permutation group is needed which would lead to a so-called nested Bethe ansatz. Therefore to be on the safe side we assume that our particles are indistinguishable, so that $\Pi_{ij}=1$ and proceed with that. Our assumption is supported by the two limiting cases, $c_{1}=0$ (a Lieb-Liniger model of interacting bosons) and $c_{0}=0$ (a Cheon-Shigehara model) which do respect this property.

%\bibitem{marquette_integrability_2013}
%I.~Marquette and J.~Links.
%\newblock {I}ntegrability of an extended $d$-wave pairing {Hamiltonian},
%\newblock \href{10.1016/j.nuclphysb.2012.09.006}{{Nucl. Phys. B}, {\bf 866}, 378 (2013)}.

%\bibitem{dukelsky_exactly_2011}
%J.~Dukelsky, S.~Lerma~H., L.~Robledo, R.~Rodriguez-Guzman, and S.~M. Rombouts,
%\newblock {E}xactly solvable pairing {Hamiltonian} for heavy nuclei,
%\newblock \href{10.1103/PhysRevC.84.061301}{Phys. Rev. C {\bf 84}, 061301 (2011)}.

%\bibitem{ortiz_what_2016}
%G.~Ortiz and E.~Cobanera,
%\newblock What is a particle-conserving {Topological} {Superfluid}? {The} fate
%  of {Majorana} modes beyond mean-field theory,
%\newblock \href{10.1016/j.aop.2016.05.020}{ Ann. Phys. {\bf 372}, 357 (2016)}.

%\bibitem{ortiz_many-body_2014}
%G.~Ortiz, J.~Dukelsky, E.~Cobanera, C.~Esebbag, and C.~Beenakker,
%\newblock {M}any-{Body} {Characterization} of {Particle}-{Conserving}
 % {Topological} {Superfluids},
%\newblock \href{10.1103/PhysRevLett.113.267002}{Phys. Rev. Lett. {\bf 113}, 267002 (2014)}.

%\bibitem{links_exact_2015}
%J.~Links, I.~Marquette, and A.~Moghaddam,
%\newblock {E}xact solution of the $p+ip$ {Hamiltonian} revisited: duality
 % relations in the hole-pair picture,
%\newblock \href{10.1088/1751-8113/48/37/374001}{J. Phys. A: Math. Theor. {\bf 48}, 374001 (2015)}.

%\bibitem{rombouts_quantum_2010}
%S.~M.~A. Rombouts, J.~Dukelsky, and G.~Ortiz,
%\newblock {Q}uantum phase diagram of the integrable $p_x+ip_y$ fermionic
 % superfluid,
%\newblock \href{10.1103/PhysRevB.82.224510}{Phys. Rev. B {\bf 82}, 224510 (2010)}.

%\bibitem{ibanez_exactly_2009}
%M.~ Iba{\~n}ez, J.~Links, G.~Sierra, and S.-Y.~Zhao,
%\newblock {E}xactly solvable pairing model for superconductors with $p_x+ip_y$-wave
%  symmetry,
%\newblock \href{10.1103/PhysRevB.79.180501}{Phys. Rev. B {\bf 79}, 180501 (2009)}.

%\bibitem{foster_quantum_2013}
%M.~S.~Foster, M.~Dzero, V.~Gurarie, and E.~A.~Yuzbashyan,
%\newblock {Q}uantum quench in a $p+ip$ superfluid: {Winding} numbers and
%  topological states far from equilibrium,
%\newblock \href{10.1103/PhysRevB.88.104511}{Phys. Rev. B {\bf 88}, 104511 (2013)}.

%\bibitem{kempkes_universalities_2016}
%S.~N. Kempkes, A.~Quelle, and C.~Morais Smith,
%\newblock Universalities of thermodynamic signatures in topological phases,
%\newblock \href{10.1038/srep38530}{Sci. Rep. {\bf 6} (2016)}.

%\bibitem{links_ground-state_2015}
%J.~Links and I.~Marquette,
%newblock {G}round-state {Bethe} root densities and quantum phase transitions,
%\newblock \href{10.1088/1751-8113/48/4/045204}{J. Phys. A: Math. Theor.,  {\bf 48},  045204 (2015)}.

%\bibitem{richardson_new_2002}
%R.~W. Richardson,
%\newblock {N}ew {Class} of {Solvable} and {Integrable} {Many}-{Body} {Models},
%\newblock \href{arXiv: cond-mat/0203512}{ arXiv:cond-mat/0203512 (2002)}.

%\bibitem{richardson_restricted_1963}
%R.~W. Richardson,
%\newblock {A} restricted class of exact eigenstates of the pairing-force  {Hamiltonian},
%\newblock \href{10.1016/0031-9163(63)90259-2}{Phys. Lett. {\bf 3}, 277 (1963)}.

%\bibitem{richardson_exact_1964}
%R.~W. Richardson and N.~Sherman,
%\newblock {E}xact eigenstates of the pairing-force {Hamiltonian},
%\newblock \href{http://dx.doi.org/10.1016/0029-5582(64)90687-X}{\em Nucl. Phys. {\bf 52}, 221 (1964)}.

\end{thebibliography}
\end{document}